# Mass Action Dynamics of Coupled Reactions using Fluctuation Theory

William R. Cannon[1] and Scott E. Baker[2]

*Author affiliation:*
[1]Biological Sciences Division and
[2]Environmental Molecular Sciences Laboratory,
Pacific Northwest National Laboratory, 902 Battelle Blvd, Richland, WA 99352.

*Corresponding author:*
William Cannon,
Fundamental and Computational Sciences Division
Pacific Northwest National Laboratory,
902 Battelle Blvd, Richland, WA 99352.
Tel: 509-375-6732.
Email: William.Cannon@pnnl.gov

**Abstract:** Comprehensive and predictive simulation of coupled reaction networks has long been a goal of biology and other fields. Currently, metabolic network models that utilize enzyme mass action kinetics have predictive power but are limited in scope and application by the fact that the determination of enzyme rate constants is laborious and low throughput. We present a statistical thermodynamic formulation of the law of mass action for coupled reactions at both steady states and non-stationary states. The formulation is based on a fluctuation theorem for coupled reactions and uses chemical potentials instead of rate constants. When used to model deterministic systems, the theory corresponds to a rescaling of the time dependent reactions in such a way that steady states can be reached on the same time scale but with significantly fewer computational steps. The significance for applications in systems biology is discussed.

## Introduction

One hundred and fifty years ago Peter Waage and Cato Maximillian Gulberg published their first article describing the law of mass action, that the rate of a chemical reaction is proportional to the concentration of the reacting species (1-4). For a simple reaction $A \underset{-1}{\overset{1}{\rightleftharpoons}} B$, the forward rate due to reaction 1 is simply,

$$\text{forward rate } = k_1[A], \quad (1)$$

where the brackets indicate the concentration the constant of proportionality $k_1$ is known as the rate constant and a similar relation exists for the reverse reaction -1. The net rate is given by,

$$\text{net rate} = k_1[A] - k_{-1}[B]. \quad (2)$$

All introductory chemistry texts describe the law of mass action in one form or another. Although the relationship is simple and can easily be applied to many reactions, the application to more complex systems such as biological metabolism is challenging because most rate constants are not available and measuring the missing rate constants is very labor intensive.

Thermodynamic (5, 6) and other approaches (7-11) to the law of mass action have been proposed that do not use rate constants, but these approaches are only valid at steady states. Studies linking thermodynamics and kinetics have historically used the concept of chemical affinities where the affinity is defined with respect to the extent of a reaction $\xi$ (12). If $\xi$ varies from 0 (no reaction) to 1 (complete stoichiometric reaction),

$$\mathcal{A} = -\frac{d\mathcal{G}}{d\xi}$$

$$\Delta\mathcal{G} = \int_0^1 \mathcal{A}(\xi)d\xi.$$

Here $G$ is the Gibbs free energy of the system. For the forward reaction +1 discussed above, the relationship between the reaction affinity and the forward $(r_1)$ and reverse $(r_{-1})$ rates at some point $\xi'$ along the reaction trajectory is,

$$\frac{r_{-1}(\xi')}{r_{+1}(\xi')} = e^{\mathcal{A}_{-1}(\xi')/RT} \quad (3)$$

where $\mathcal{A}_{-1}$ is the affinity of the reaction in the reverse direction. In this case, the net rate of a reaction is (13),

$$\text{net rate} = r_1\left(1 - e^{\mathcal{A}_{-1}/RT}\right).$$

Consequently, the usual kinetic description of the reaction rate can be forumulated to include a thermodynamic term when a reaction doesn't change the abundance of the reactants or products significantly. In this case,

$$\text{net rate} \approx r_1\left(1 - e^{-\Delta\mathcal{G}_{-1}/RT}\right) \quad (4)$$

For deterministic systems in which the concentrations do not change (steady state), $\mathcal{A}(\xi)$ is constant and $\Delta\xi = 1$. Consequently, Eqn 4 is exact, as suggested by Temkin (14), but only for macroscopic and deterministic steady state systems. To see that this is the case, one only need to consider the initial phase of a reaction when only reactants for the forward reaction are present. In this case, Eqn 4 results in a finite reverse reaction rate because $\Delta\mathcal{G}_{-1} = -\Delta\mathcal{G}_1$ and $\Delta\mathcal{G}_1$ has a finite value. One could argue that if the reactants for the reverse reaction are at zero the exponent of the free energy change for the reverse reaction is zero ($\Delta\mathcal{G}_{-1}$ has an infinitely large positive value). However, the logic of this argument results in the contradiction that $\Delta\mathcal{G}_{-1} \neq -\Delta\mathcal{G}_1$ and therefore free energy is no longer a function of state. Even at steady state, when the number of particles of the reaction intermediates drop below $\sim 10$, Eqn 4 will not produce the correct results. A single firing of a reaction in this case can easily result in changes in the abundance of chemical species of 10-100%. This discussion suggests that a microscopic function related to the free energy change is needed.

In this regard, thermodynamic approaches to microscopic dynamics have used statistical fluctuation theories (15-18). Fluctuation equations are developed using stochastic models. The usual differential equation relating rate to concentrations (Eqn. 2) is replaced by the change in probability of a state as a function of time. For a chemical system, a microscopic state may be defined as the set of all counts or concentrations of each chemical species. If one chemical species increases by just one count, then the system moves to a different state. If the system is in some state *J* at time *t*, then it may move into another state *K* due to an increase or decrease in the count of some chemical species due to a reaction. Or, if some other reaction occurred, the system would move from state *J* into a different state *L*.

Fluctuation theorems can relate the probabilities of conjugate processes *from the same original state.* If $\pi(1\,|\,J)$ is the probability of a transition from state *J* to another state through process 1 and $\pi(-1\,|\,J)$ is the probability of a transition from state *J* through the conjugate process -1, the ratio of these probabilities is related to the difference in how much energy is dissipated through each process,

$$\frac{\pi(1\,|\,J)}{\pi(-1\,|\,J)} = e^{\Delta\Omega_{1,-1}(J)}, \qquad 5$$

where $\Delta\Omega_{1,-1}(J)$ is a measure of the difference in the dissipation of energy between processes 1 and -1 if both processes start from state $J$. This is a significant aspect of fluctuation theorems – they relate the ratio of the probabilities of conjugate processes to a value that is potentially measurable or calculable. Not coincidentally, Eqn 5 looks eerily similar in form to Eqn 3; in fact, Eqn 5 is the stochastic version of Eqn 3, where the concept of the discrete microstate is replaced by the continuous parameter $\xi'$ and the reaction affinity $\mathcal{A}_1(\xi'=0)$ maps to $\Delta\Omega_{1,-1}(J)$. If one could determine $\Delta\Omega_{1,-1}(J)$ then the relative reaction probabilities could be determined and non-steady state systems could be modeled without the use of the rate constants first described by Waage and Gulberg over 150 years ago.

In earlier reports, we used a statistical thermodynamics formulation of reactions (19) which used fluctuation theorems to compare the dynamics of microstate transitions of different versions of the tricarboxylic acid cycles (20) found in organisms occupying different ecological niches. The models based in the latter work did not use rate constants but instead used an assumption of maximum entropy to model the thermodynamically optimal dynamics of the systems. Earlier, Gaspard introduced a fluctuation theory for the mass action dynamics of chemical reactions (21), but formulated $\Delta\Omega_{1,-1}(J)$ using rate constants. In this report, we demonstrate the relationship between kinetic simulations using rate constants and earlier work (19, 20, 22) using statistical thermodynamics. In doing so, a stochastic theory for coupled reactions is presented that formulates a fluctuation theory approach based on chemical potentials which can provide exact rates for all reactions for which steady state levels of reaction intermediates are measurable, obviating the need for rate constants. At macroscopic scale, this fluctuation theorem is shown to be the rescaled deterministic rate law.

## Theory

*A fluctuation theory for isolated reactions***.** Fluctuation equations are developed using stochastic models (17). In stochastic models of chemical reactions (Markov models), the usual differential equation relating rate to concentrations is replaced by the change in probability of a state as a function of time. For a chemical system, a microscopic state may be defined as the set of all counts or concentrations of each chemical species. If one chemical species increases by just one count, then the system moves to a different state as shown in Figure 1. If the system is in some state *K* at time *t*, then it may move into another state *J* due to an increase or decrease in the count of some chemical species due to a reaction that occurs as the time increases by $\delta t$ to $t+\delta t$. Likewise, if the system is already in state *J* at time *t*, then it may move out to another state due to some other reaction that occurs in time $\delta t$. The change in probability of state *J* at time *t* in time due to an increase in time by $\delta t$, $\delta \mathcal{P}r(J \mid \delta t, t)$, is then,

$$\begin{aligned}\delta \mathcal{P}r(J \mid \delta t,t) &= \delta \mathcal{P}r^{in}(J \mid \delta t,t) - \delta \mathcal{P}r^{out}(J \mid \delta t,t)\\ &= \sum_{\text{states } K} \mathcal{P}r(K \mid t)\pi(J \mid K,\delta t) - \sum_{\text{states } K} \mathcal{P}r(J \mid t)\pi(K \mid J,\delta t),\end{aligned}$$

where $\pi(J \mid K,\delta t)$ and $\pi(K \mid J,\delta t)$ are transition probabilities from state $K$ to $J$ and $J$ to $K$, respectively, due to an increase in time $\delta t$. $\mathcal{P}r(J \mid t)$ is the probability of state $J$ at time $t$. If a state transition can only occur through a chemical reaction, then the state-to-state transition probabilities are reaction probabilities as well. Expressing the transition probabilities as reaction probabilities allows direct comparison to the law of mass action for chemical reactions (Eqn. 2), as discussed below. In this case, if the transition from state *J* to state *K* occurs due to reaction 1, as shown in Figure 1, then $\pi(K \mid J,\delta t)$ is the reaction probability $\pi(1 \mid J,\delta t)$ in which the argument 1 indicates that the transition from *J* is due to reaction 1.

Fluctuation theories are based on comparing the probability of a forward trajectory $\pi(1 \mid \delta t)$ to the probability of the reverse trajectory $\pi(-1 \mid \delta t)$ using a statistical odds. Excellent reviews of fluctuation theorems are available elsewhere (17). In the context of chemical reactions, fluctuation theories can be most easily understood by analogy to the equilibrium situation for a stochastic reaction, in which the condition of detailed balance exists. In this case, the forward reaction is just as likely as the reverse reaction. For systems that obey detailed balance and have thermal energy, it is actually the weighted average of microstates for which detailed balance exists. For the simple system $A \underset{-1}{\overset{1}{\rightleftharpoons}} B$, if we denote an reactant state as *K* and a product state as *J* then detailed balance is such that,

$$\sum_{\text{states } K} \mathcal{Pr}(K \mid t)\pi(1 \mid K,\delta t) = \sum_{\text{states } J} \mathcal{Pr}(J \mid t)\pi(-1 \mid J,\delta t),$$

and hence the value of the ratio of the probabilities of the forward and reverse processes averaged over many trajectories is unity,

$$\frac{\sum_{\text{states } K} \mathcal{Pr}(K \mid t)\pi(1 \mid K,\delta t)}{\sum_{\text{states } J} \mathcal{Pr}(J \mid t)\pi(-1 \mid J,\delta t)} = 1$$

$$\left\langle \frac{\pi(1 \mid K,\delta t)}{\pi(-1 \mid J,\delta t)} \right\rangle = 1$$

where the brackets $\langle \ \rangle$ indicate an average over all states. In fact, the designation of one state as a reactant state (*K*) and another as a product state (*J*) is arbitrary as a product state can also represent a reactant state, as shown in Figure 1. In a microscopic system containing thermal energy, it may actually be the time average of the above quantities that constitute equilibrium. At any instant in time, the system may experience a small fluctuation away from the detailed balance condition, but over a longer time the fluctuations cancel out.

In a non-equilibrium situation, the average value of the ratio of the probabilities of a process is some value other than 1, indicating which is most likely, the forward process or the reverse process. The average value is related to the entropy of a change of state, $\Delta S_{-1,1}$,

$$\left\langle \frac{\pi(1 \mid K,\delta t)}{\pi(-1 \mid J,\delta t)} \right\rangle = e^{\Delta S_{-1,1}}, \qquad 6$$

where the brackets again $\langle \ \rangle$ indicate an macroscopic average over all microscopic states $J,K$. This relation is analogous to Crook's fluctuation theorem applied to the context of chemical reactions (15).

A fluctuation theorem may have a form similar to Eqn. 6 or it may have several important differences. First, the average value need not always be used. When the average value is not used the actual value of the exponent is determined by the amount of energy that is dissipated from the system, $\Omega$. Importantly, fluctuation theorems can relate conjugate processes originating from the same initial state via the dissipation,

$$\begin{aligned} \frac{\mathcal{Pr}(J \mid t)\pi(1 \mid J,\delta t)}{\mathcal{Pr}(J \mid t)\pi(-1 \mid J,\delta t)} &= e^{\Delta\Omega_{1,-1}(J)} \\ \frac{\pi(1 \mid J,\delta t)}{\pi(-1 \mid J,\delta t)} &= e^{\Delta\Omega_{1,-1}(J)}, \end{aligned} \qquad 7$$

where $\Delta\Omega_{1,-1}(J)$ is the difference in the dissipation for processes 1 and -1 from state $J$. From a statistical standpoint, the odds ratio above is simply the ratio of the likelihood of observing one product state to another after a time $\delta t$. The likelihood for a state with $n_j$ particles of type *j* and $N_{total}$ objects is (19),

$$\Pr(n_1,...,n_m \mid N_{total},\theta_1,...,\theta_m) = N_{total}! \prod_{\text{objects } j}^{m} \frac{1}{n_j!}\theta_j^{n_j}. \qquad 8$$

where $\theta_j$ is the probability of an object being of type *j.* Each different distribution of the $N_{total}$ objects among the *m* different types of objects is a different microscopic state and is characterized by the probability density of Eqn. 8. For a system with the total number of particles $N_{total}$ constant, the odds of a transition probability from state *J* to state *K* via reaction 1 relative to a transition from state *J* to state *L* is,

$$\begin{aligned}\frac{\pi(1\mid J)}{\pi(-1\mid J)} &= \frac{\Pr(n_1(K),...,n_m(K)\mid N_{total},\theta_1,...,\theta_m)}{\Pr(n_1(L),...,n_m(L)\mid N_{total},\theta_1,...,\theta_m)} \\ &= \prod_{\text{species } j} \frac{n_j(L)!}{n_j(K)!}\theta_j^{n_j(K)-n_j(L)}\end{aligned} \qquad 9$$

where $\Pr(n_1(K),...,n_m(K)\mid N_{total}(K),\theta_1,...,\theta_m)$ is the system probability density for state *J.* In order to look at the odds ratio as a function of the counts for state *J,* the identity $n_j(K) = n_j(J)+\gamma_j$ where $\gamma_j$ is the stoichiometric coefficient for species *j* in reaction 1 is used such that

$$\begin{aligned}\prod_{\text{species } j} \frac{n_j(L)!}{n_j(K)!} &= \frac{\prod_{\substack{\text{reactant}\\ \text{species } j}}^{\text{Rxn 1}} \left(n_j(L)\right)_{v_j}}{\prod_{\substack{\text{reactant}\\ \text{species } j}}^{\text{Rxn -1}} \left(n_j(K)\right)_{v_j}} \\ &= \frac{\prod_{\substack{\text{product}\\ \text{species } j}}^{\text{Rxn -1}} n_j(J)^{(v_j+1)}}{\prod_{\substack{\textit{product}\\ \text{species } j}}^{\text{Rxn 1}} n_j(J)^{(v_j+1)}}\end{aligned}$$

where $n_j(J)+\gamma_j = n_j(K)$. The function $\left(n_j\right)_{\gamma_j} = n_j\cdot\left(n_j-1\right)\cdots\left(n_j-\gamma_j+1\right)$ is the falling factorial function and likewise $n_j^{(\gamma_j+1)} = n_j\cdot\left(n_j+1\right)\cdots\left(n_j+\gamma_j\right)$ is the rising factorial function. The last equality above makes it clear that Eqn. 9 is the odds of two product states that originate from the same reactant state *J*.

If the system is such that the total number of objects varies, then the probability densities for two different states *L* and *K* may be such that in the ratio of the probability densities the $N_{total}!$ terms of Eqn 8 do not cancel out. The probability densities are for systems of different sizes. The odds ratio is normalized by taking into account the full extent of sampling space for each possible distribution. In this case,

$$\frac{\pi(1|J)}{\pi(-1|J)} = q^{\Delta N_{total}(K,L)} \cdot \prod_{\text{species } j} \frac{n_j(L)!}{n_j(K)!} \theta_j^{\,n_j(K)-n_j(L)}$$

Recognizing that $q^{\Delta N_{total}(K,L)} = e^{-\Delta N_{total}(K,L)\cdot\mu^0/RT}$ where $\mu^0$ is the standard chemical potential, the first quantity on the left hand side is simply the odds of adding $\Delta N_{total}(K,L)$ particles to the open system. When the total number of particles does not vary, $q^{\Delta N_{total}(K,L)} = 1$. In this case, as long as the probabilities $\theta_1,\ldots,\theta_m$ do not vary with time and there is a reaction allowing for a transition between two states the transition probability can be written as the product of a time-independent constant of the reaction and the time-dependent counts,

$$\frac{\pi(1|J)}{\pi(-1|J)} = \left( \prod_{\text{species } j} \theta_j^{\,n_j(K)-n_j(L)} \right) \cdot \prod_{\text{species } j} \frac{n_j(L)!}{n_j(K)!} = \mathcal{K}_{\pm 1} \cdot \frac{\prod_{\substack{\text{reactant} \\ \text{species } j}}^{\text{Rxn 1}} \left(n_j(L)\right)_{v_j}}{\prod_{\substack{\text{reactant} \\ \text{species } j}}^{\text{Rxn -1}} \left(n_j(K)\right)_{v_j}} \qquad 10$$

Here the constant $\mathcal{K}_{\pm 1}$ is proportional to the ratio of the probabilities of the species involved in the reaction – it is the equilibrium constant. As long as the probabilities do not change with time, $\mathcal{K}_{\pm 1}$ is the same for any state-to-state transitions involving reactions 1 and -1. For large $n_A$ and small $\gamma_A$ the factorials can be approximated by powers,

$$\frac{\pi(1|J)}{\pi(-1|J)} \approx \mathcal{K}_{\pm 1} \frac{\prod_{\substack{\text{reactant} \\ \text{species } j}}^{\text{Rxn 1}} n_j(L)^{v_j}}{\prod_{\substack{\text{reactant} \\ \text{species } j}}^{\text{Rxn -1}} n_j(K)^{v_j}} \qquad 11$$

From comparison of Eqn. 10 and Eqn. 7 it is clear that when the total number of particles do not change,

$$\Delta\Omega_{1,-1}(J) = \log\left( \mathcal{K}_{\pm 1} \cdot \frac{\prod_{\substack{\text{reactant} \\ \text{species } j}}^{\text{Rxn 1}} \left(n_j(L)\right)_{v_j}}{\prod_{\substack{\text{reactant} \\ \text{species } j}}^{\text{Rxn -1}} \left(n_j(K)\right)_{v_j}} \right) \qquad 12$$

Although the quantity in parentheses may look like the product of the equilibrium constant and the reaction quotient – and thus related to the change in macroscopic free energy and entropy – this is only approximately true for larger systems. The macroscopic free energy and entropy are averages over many microscopic states while Eqn. 12 is not an average. Moreover, the free energy and entropy change of a reaction pertain to the odds of the current state relative to the next state. In contrast, Eqn. 10 and Eqn. 12 are similar to other fluctuation theorems in that they pertain to the odds of reaching different product states from the current state. The differences between the fluctuation theorem of Eq. 10 and the macroscopic free energy change are important in that the dynamics will not be correct if improperly implemented. This distinction is the reason why Eqn 4 is correct only for special cases – steady state systems with a large number of particles.

For isolated reactions this framework can completely describe the dynamics. To model coupled or sequential reactions $1,2,\ldots,v$ a similar fluctuation theorem is needed for the odds of one reaction relative to the next reaction, $\pi(i+1|J)/\pi(i|J)$.

For two reactions that, instead of being opposite reaction paths (e.g., Figure 1: $L\xleftarrow{-1}J\xrightarrow{1}K$), are instead unrelated (e.g., Figure 1: $K\xleftarrow{1}J\xrightarrow{2}M$), a dissipation function analogous to Eqn 12 must be found. The odds of reaction 1 (from state *J* to state *K* following the scheme in Figure 1) to reaction 2 (from state J to state M) are,

$$\begin{aligned}\frac{\pi(2|J)}{\pi(1|J)} &= \left(q^{\Delta N_{total}(M,K)}\cdot\prod_{\text{species } j}\theta_j^{n_j(M)-n_j(K)}\right)\cdot\prod_{\text{species } j}\frac{n_j(K)!}{n_j(M)!}\\ &= q^{\Delta N_{total}(M,K)}\cdot\mathcal{K}_{2,1}\cdot\frac{\prod_{\substack{\text{reactant}\\\text{species } j}}^{\text{Rxn 2}}\left(n_j(J)\right)_{v_j}}{\prod_{\substack{\text{reactant}\\\text{species } j}}^{\text{Rxn 1}}\left(n_j(J)\right)_{v_j}}\end{aligned} \qquad 13$$

The challenge in applying Eqn 13 is that $\mathcal{K}_{2,1}$ — the ratio of the product of the probabilities of the chemical products of different reactions in a system of coupled reactions — cannot be determined from equilibrium measurements or calculated from free energies of formation. Like rate constants, coupling constants such as $\mathcal{K}_{2,1}$ are system dependent and have no system-independent standard value. Next, we demonstrate how non-equilibrium steady state measurements can be used to determine coupling constants and how coupled reactions can be modeled without rate constants.

*Coupling Between Reactions.* For a simple coupled reaction system,

$$\gamma_A A \underset{-1}{\overset{1}{\rightleftharpoons}} \gamma_{\mathbf{B}} B \underset{-2}{\overset{2}{\rightleftharpoons}} \gamma_C C \qquad \text{Scheme A}$$

where $A$, $B$ and $C$ are chemical species and $\gamma_A, \gamma_B$ and $\gamma_C$ are the stoichiometric coefficients for the reactions, the states of the system are composed of the counts or concentrations of each of the chemical species $A, B$ and $C$. For example, a state *J* consists of $N_A(J)$ counts of chemical *A*, $N_B(J)$ of chemical *B*, and $N_C(J)$ of chemical *C*. The probability of any state *J* at time *t* is denoted $\mathcal{Pr}(J|t)$. A different state *K* is reached from *J* due to a change of state $\delta S$ such that $K = J + \delta S$. The probability of such a change from state *J* to state *K* in a time interval $\delta t$ is $\pi(K|J,\delta t)$. A Markov model that describes changes of state is given by,

$$\mathcal{Pr}(J \mid t+\delta t) = \mathcal{Pr}(J \mid t) + \delta\mathcal{Pr}(J \mid t,\delta t)$$

$$= \mathcal{Pr}(J \mid t) + \sum_{states\ \mathrm{K}} \mathcal{Pr}(K \mid t)\pi(J \mid K,\delta t) - \mathcal{Pr}(J \mid t)\pi(K \mid J,\delta t)$$

For the system in Scheme A with accessible states shown graphically in Figure 1, the probability flux out of state *J* is given by,

$$\delta\mathcal{Pr}^{out}(J \mid t,\delta t) = \mathcal{Pr}(J \mid t)\pi(K \mid J,\delta t) + \mathcal{Pr}(J \mid t)\pi(L \mid J,\delta t) + \mathcal{Pr}(J \mid t)\pi(M \mid J,\delta t) + \mathcal{Pr}(J \mid t)\pi(N \mid J,\delta t) \qquad 14$$

where *K*, *L*, *M* and *N* are the states reached from state *J* through reactions 1, -1, 2 and -2, respectively (Figure 1). As above, we can represent the state-to-state transition probabilities $\pi(K \mid J,\delta t)$ as reaction probabilities $\pi(1 \mid J,\delta t)$ in which the argument 1 indicates that the transition from *J* to *K* is due to reaction 1. Rearranging terms in Eqn. 14,

$$\frac{\delta\mathcal{Pr}^{out}(J \mid t,\delta t)}{\mathcal{Pr}(J \mid t)\cdot\pi(1 \mid J,\delta t)} = \left(1 + \frac{\pi(-1 \mid J,\delta t)}{\pi(1 \mid J,\delta t)}\right) + \frac{\pi(2 \mid J,\delta t)}{\pi(1 \mid J,\delta t)}\left(1 + \frac{\pi(-2 \mid J,\delta t)}{\pi_2(2 \mid J,\delta t)}\right). \qquad 15$$

Each term other than the constants in Eqn. 15 is a fluctuation theorem relationship. Eqn. 15 is significant in that it tells us how to model transitions from any state *J* to any other state accessible through a set of reactions using the odds ratios. In other words, the transitions can be modeled using the equilibrium constants or chemical potentials instead of rate constants. Next, a microscopic rate law is derived that allows one to determine $\pi(2 \mid J)/\pi(1 \mid J)$.

*A Microscopic Rate Law*. The flux into state *J* can only occur through these same four states shown in Figure 1 such that,

$$\begin{aligned}\delta\mathcal{Pr}^{in}(J \mid t,\delta t) &= \mathcal{Pr}(L \mid t)\cdot\pi(J \mid L,\delta t) + \mathcal{Pr}(K \mid t)\cdot\pi(J \mid K,\delta t) + \\ &\quad \mathcal{Pr}(N \mid t)\cdot\pi(J \mid N,\delta t) + \mathcal{Pr}(M \mid t)\cdot\pi(J \mid M,\delta t) \\ &= \mathcal{Pr}(L \mid t)\cdot\pi(1 \mid L,\delta t) + \mathcal{Pr}(K \mid t)\cdot\pi(-1 \mid K,\delta t) + \\ &\quad \mathcal{Pr}(N \mid t)\cdot\pi(2 \mid N,\delta t) + \mathcal{Pr}(M \mid t)\cdot\pi(-2 \mid M,\delta t)\end{aligned}$$

Grouping the terms by each respective reaction, the total probability flux through state *J* is then,

$$\begin{aligned}\delta\mathcal{Pr}(J \mid t,\delta t) &= \delta\mathcal{Pr}^{in}(J \mid t,\delta t) - \delta\mathcal{Pr}^{out}(J \mid t,\delta t) \\ &= \left(\mathcal{Pr}(L \mid t)\pi(1 \mid L,\delta t) - \mathcal{Pr}(J \mid t)\pi(1 \mid J,\delta t)\right) + \\ &\quad \left(\mathcal{Pr}(K \mid t)\pi(-1 \mid K,\delta t) - \mathcal{Pr}(J \mid t)\pi(-1 \mid J,\delta t)\right) + \\ &\quad \left(\mathcal{Pr}(N \mid t)\pi(2 \mid N,\delta t) - \mathcal{Pr}(J \mid t)\pi(2 \mid J,\delta t)\right) + \\ &\quad \left(\mathcal{Pr}(M \mid t)\pi(-2 \mid M,\delta t) - \mathcal{Pr}(J \mid t)\pi(-2 \mid J,\delta t)\right)\end{aligned}$$

The notation can be simplified to emphasize the reaction probabilities by using $\pi_1(\cdot)$ to indicate that the transition is due to reaction 1 and the dot to indicate that the value is a function of the originating state. Likewise $\pi_{-1}(\cdot),\pi_2(\cdot)$, and $\pi_{-2}(\cdot)$ indicate that the transition is due to reactions, -1, 2 and -2, respectively, and are functions of the originating state. Using this notation,

$$\delta\mathcal{P}r(J\,|\,t,\delta t) = \left(\mathcal{P}r(L\,|\,t) - \mathcal{P}r(J\,|\,t)\right)\pi_1(\cdot) + \left(\mathcal{P}r(K\,|\,t) - \mathcal{P}r(J\,|\,t)\right)\pi_{-1}(\cdot) + \\ \left(\mathcal{P}r(N\,|\,t) - \mathcal{P}r(J\,|\,t)\right)\pi_2(\cdot) + \left(\mathcal{P}r(M\,|\,t) - \mathcal{P}r(J\,|\,t)\right)\pi_{-2}(\cdot).$$

At equilibrium each pair of forward and reverse reactions are equally likely - the reaction system has symmetry. If a non-equilibrium system is such that the net flow of material proceeds from some initial state *L* or *N* to intermediate state *J* and then to a final state *K* or *M* (left to right in Figure 1), then

$$\mathcal{P}r(K\,|\,t), \mathcal{P}r(M\,|\,t) \ge \mathcal{P}r(J\,|\,t) \ge \mathcal{P}r(L\,|\,t), \mathcal{P}r(N\,|\,t).$$

Rearranging the terms so that the probability difference in parentheses (corresponding to the same reaction) is positive,

$$\delta\mathcal{P}r(J) = -\left(\mathcal{P}r(J\,|\,t) - \mathcal{P}r(L\,|\,t)\right)\pi_1(\cdot) + \left(\mathcal{P}r(K\,|\,t) - \mathcal{P}r(J\,|\,t)\right)\pi_{-1}(\cdot) + \\ \left(\mathcal{P}r(N\,|\,t) - \mathcal{P}r(J\,|\,t)\right)\pi_2(\cdot) - \left(\mathcal{P}r(J\,|\,t) - \mathcal{P}r(M\,|\,t)\right)\pi_{-2}(\cdot)$$

Here $\left(\mathcal{P}r(J\,|\,t) - \mathcal{P}r(L\,|\,t)\right)\pi_1(\cdot)$ is the instantaneous probability flux through reaction 1 and likewise for the other terms above. An equation analogous to the macroscopic differential equation for mass action kinetics (rate law) is obtained by averaging the probability flux over all states,

$$\sum_J \delta\mathcal{P}r(J\,|\,\delta t,t) = \sum_{J,K,L,M,N}^{all} -\left(\mathcal{P}r(J\,|\,t) - \mathcal{P}r(L\,|\,t)\right)\pi_1(\cdot) + \left(\mathcal{P}r(K\,|\,t) - \mathcal{P}r(J\,|\,t)\right)\pi_{-1}(\cdot) \\ + \left(\mathcal{P}r(N\,|\,t) - \mathcal{P}r(J\,|\,t)\right)\pi_2(\cdot) - \left(\mathcal{P}r(J\,|\,t) - \mathcal{P}r(M\,|\,t)\right)\pi_{-2}(\cdot)$$

$$\langle\delta\mathcal{P}r\rangle = \mp\langle\pi_1\rangle \pm \langle\pi_{-1}\rangle \pm \langle\pi_2\rangle \mp \langle\pi_{-2}\rangle,$$

Whether a reaction has a positive or negative coefficient depends on the net flow of materials through each of the coupled reactions in the system. If there is no net flow because the system is at equilibrium, then the difference in the state probabilities for each parenthetical pair averages out to zero. Non-equilibrium conditions break the symmetry of the forward and reverse reactions such that the reactions are no longer equally likely:

$$\text{if } \langle\delta\mathrm{Pr}\rangle \ge 0, \text{ then } \langle\delta\mathrm{Pr}\rangle = -\langle\pi_1\rangle + \langle\pi_{-1}\rangle + \langle\pi_2\rangle - \langle\pi_{-2}\rangle,$$
$$\text{if } \langle\delta\mathrm{Pr}\rangle \le 0, \text{ then } \langle\delta\mathrm{Pr}\rangle = \langle\pi_1\rangle - \langle\pi_{-1}\rangle - \langle\pi_2\rangle + \langle\pi_{-2}\rangle.$$

Regardless of the direction of the non-equilibrium forces, the change in probability of the system can be renormalized such that,

$$\frac{\mp\langle\delta\mathcal{P}r\rangle}{\langle\pi_1\rangle} = \left(1 - \frac{\langle\pi_{-1}\rangle}{\langle\pi_1\rangle}\right) - \frac{\langle\pi_2\rangle}{\langle\pi_1\rangle}\left(1 - \frac{\langle\pi_{-2}\rangle}{\langle\pi_2\rangle}\right). \qquad 16$$

At steady state, the average change in probability of the states $\langle\delta\mathcal{P}r\rangle = 0$. This leads to the steady state (*SS*) coupling term,

$$\frac{\langle\pi_2\rangle_{SS}}{\langle\pi_1\rangle_{SS}} = \left(1 - \frac{\langle\pi_{-1}\rangle_{SS}}{\langle\pi_1\rangle_{SS}}\right) \Big/ \left(1 - \frac{\langle\pi_{-2}\rangle_{SS}}{\langle\pi_2\rangle_{SS}}\right). \qquad 17$$

The value of the coupling term for the reactions in Scheme A at any state $J$ is then,

$$\begin{aligned}\frac{\pi(2\,|\,J,\delta t)}{\pi(1\,|\,J,\delta t)} &= \frac{\langle\pi_2\rangle_{SS}}{\langle\pi_1\rangle_{SS}}\frac{n_A(SS)^{(\gamma_A)}}{n_B(SS)^{(\gamma_B)}}\frac{n_B(J)^{(\gamma_B)}}{n_A(J)^{(\gamma_A)}} \\ &= \mathcal{K}_{2,1}\frac{n_B(J)^{(\gamma_B)}}{n_A(J)^{(\gamma_A)}} \\ &\approx \mathcal{K}_{2,1}\frac{n_B{}^{\gamma_B}}{n_A{}^{\gamma_A}}\end{aligned} \qquad 18$$

where $n_A{}^{(\gamma_A)} = n_A\cdot(n_A+1)\cdots(n_A+\gamma_A-1)$ is again the rising factorial function and the approximation holds for large $n_A$ and small $\gamma_A$.

At the macroscopic limit of large numbers of each chemical species, the time dependent probability of a state is represented by the time dependent probability of the concentrations of the chemical species themselves. In this case, the difference equation 16 is replaced by a differential equation that represents the mass action rate law. The probability of reaction 1 in Scheme A can be associated with a rate constant $k_1$ such that $\langle\pi_1\rangle = k_1[A]/c_{norm}$ where $[A]$ is the concentration of species $A$ and $c_{norm}$ is a normalization constant. Analogous relationships hold for the other reaction probabilities such that Eqn 16 becomes,

$$\begin{aligned}\frac{c_{norm}}{k_1[A]}\cdot\frac{d[B]}{dt} &= \left(1-\frac{k_{-1}[B]}{k_1[A]}\right)-\frac{k_2[B]}{k_1[A]}\left(1-\frac{k_{-2}[C]}{k_2[B]}\right), \\ &= \left(1-\mathcal{K}_{-1}\frac{[B]}{[A]}\right)-\mathcal{K}_{2,1}\frac{[B]}{[A]}\left(1-\mathcal{K}_{-2}\frac{[C]}{[B]}\right).\end{aligned}$$

Thus, at large numbers of chemical species the microscopic rate law equation (Eqn 16) is analogous to a rescaling of the usual deterministic mass action rate law.

**Results and Discussion**

In biology and many other fields reactions rarely occur in isolation. Instead, multiple reactions may be coupled together that transform initial reactants into final products. Consider the simple coupled reaction system,

$$\gamma_A A \underset{-1}{\overset{1}{\rightleftharpoons}} \gamma_{\mathbf{B}} B \underset{-2}{\overset{2}{\rightleftharpoons}} \gamma_C C \qquad \text{Scheme A}$$

where $A$, $B$ and $C$ are chemical species and $\gamma_A, \gamma_B$ and $\gamma_C$ are the stoichiometric coefficients for the reactions.

Coupled reaction theory has been developed, as other fluctuation theories, using a probabilistic approach. Because of the law of large numbers, the deterministic formulation of coupled reaction theory is quite straightforward. In this case, coupled reaction theory can be thought of as a time rescaling of the dynamical equations. For reaction scheme show in Scheme A with $\gamma_A = \gamma_B = \gamma_C = 1$, the deterministic time dependence of each molecule is governed by the usual set of ordinary differential equation. For species $B$ this time dependence is,

$$\frac{d[B]}{dt} = k_1[A] - k_{-1}[B] - k_2[B] + k_{-2}[C]. \tag{19}$$

In coupled reaction theory, the analogous deterministic rate equation is rescaled in time,

$$\tau_1 \cdot \frac{d[B]}{dt} = \left(1 - \mathcal{K}_{-1}\frac{[B]}{[A]}\right) - c_{21}\left(1 - \mathcal{K}_{-2}\frac{[C]}{[B]}\right), \tag{20}$$

Here, we refer to $\tau_1 = \left(k_1[A]\right)^{-1}$ as the characteristic time for reaction 1 when the concentration of the reactants is [A]. $\mathcal{K}_{-1} = k_{-1}/k_1$ and $\mathcal{K}_{-2} = k_{-2}/k_2$ are equilibrium constants. Although $\tau_1$ and $d[B]/dt$ cannot be calculated separately without knowledge of rate constants, their product can be determined. The second term in each parenthetical expression corresponds to a fluctuation theory in stochastic systems. Although these terms look like the usual free energies, this is only true in the limit of a continuous system in which the difference between states is infinitesimal. In the limit that each state is discrete, these terms correspond to Eqn 5 where counts are used instead of concentrations,

$$\begin{aligned}\Delta\Omega_{1,-1}(J) &\approx \log \mathcal{K}_{-1}\frac{\left(B(J)+\gamma_{\mathbf{B}}\right)^{\gamma_{\mathbf{B}}}}{\left(A(J)+\gamma_A\right)^{\gamma_A}} \\ &= \log \mathcal{K}_{-1}\frac{B(K)^{\gamma_{\mathbf{B}}}}{A(L)^{\gamma_A}}\end{aligned} \tag{21}$$

In contrast, the free energy is the log of the macroscopic average value of the likelihood for reaction -1. The likelihood is the ratio of the probability density of the reactant state *J* to the product state *L* relative to the equilibrium state,

$$\begin{aligned}LR_1 &= \mathcal{K}_{-1}\frac{B(J)^{\gamma_{\mathbf{B}}}}{\left(A(J)+\gamma_A\right)^{\gamma_A}} \\ &= \mathcal{K}_{-1}\frac{B(J)^{\gamma_{\mathbf{B}}}}{A(L)^{\gamma_A}}\end{aligned}$$

Finally, the free energy change is the log of the macroscopic average value of likelihood,

$$\Delta\mathcal{G}_{-1} = -RT\log\mathcal{K}_{-1}\left\langle\frac{B(J)^{\gamma_{\mathbf{B}}}}{A(L)^{\gamma_A}}\right\rangle$$

where *T* is the absolute temperature and *R* is the gas constant and the brackets indicate a concentration, which is an average value of the counts divided by a volume. (Equilibrium constants, in contrast to rate constants, are unitless.) Consequently, the use of a free energy change in Eqn 4 for systems is only valid for macroscopic steady state systems in which the system is large enough that deviations from the average are insignificant.

In a stochastic simulation, the choice of whether a forward reaction 1 is chosen over a reverse reaction -1 from a state *J* depends on the likelihood ratio in Eqn 21. The choice of whether reaction 2 is chosen over reaction 1, however, depends on the coupling term, $c_{21}$. The coupling term $c_{21} = k_2[B]/k_1[A]$ is determined by solving Eqn 20 at steady state,

$$c_{21}^{SS} = \left(1 - \mathcal{K}_{-1}\frac{[B]}{[A]}\right) \Big/ \left(1 - \mathcal{K}_{-2}\frac{[C]}{[B]}\right). \qquad 22$$

*Equations 20 and 22 tells us that if we measure concentrations of metabolites at a steady state, we can determine the relative dynamics of a system without rate constants.* The parameters that are needed are the steady state concentrations and equilibrium constants (or standard free energies of reaction). With these parameters in hand, it is easy to determine the non-steady state value of the coupling term. For the reactions in Scheme A with $\gamma_A = \gamma_B = \gamma_C = 1$ and steady state concentrations denoted by the subscript *SS*,

$$\begin{aligned} c_{21} &= c_{21}^{SS} \cdot \frac{[A]_{SS}}{[B]_{SS}} \cdot \frac{[B]}{[A]} \\ &= \mathcal{K}_{21} \cdot \frac{[B]}{[A]}. \end{aligned} \qquad 23$$

where $\mathcal{K}_{21}$ is a constant representing the ratio of the chemical potentials of reactants for reactions 2 and 1. Like equilibrium constants, for elementary reactions coupling constants are the ratio of the rate constants of the respective reactions. Unlike equilibrium constants, however, coupling constants such as $\mathcal{K}_{21}$ are system dependent. In analogy to other fluctuation theorems, we refer to this approach as coupled reaction theory. For the reactions of Scheme A, Figure 2 shows the comparison between a simulation solving the stochastic equivalent of Eqn 19 and one solving for stochastic equivalent of Eqn 20. In these simulations, the absolute time is not modeled – all results pertain to the characteristic time of the +1 reaction in Scheme A. However, if just one rate constant is available then coupled reaction theory can reproduce the exact time-dependent trajectory as the trajectory governed by the (un-scaled) ordinary differential equation Eqn 19.

The top row in Figure 2 compares the steady state trajectories of the reaction intermediate *B* from stochastic simulations using coupled reaction theory with trajectories from stochastic kinetic simulations (the steady state solutions of the deterministic ordinary differential equation are shown for comparison, also). When using the same set of random numbers, the trajectories from the coupled reaction theory simulations are exactly the same as that for the stochastic kinetic simulations (the trajectories are offset by +5 counts for clarity). We can also demonstrate that the reaction probabilities are identical.

Coupled reaction theory is applicable away from the steady state, as well. Shown in row 2 in Figure 2 are the transient decays from non-equilibrium states to the steady states. As expected, both the stochastic kinetic simulations and the coupled reaction theory simulations produce exactly the same trajectory when using the same set of random numbers. In the case of rows 1-2 in Figure 2, the forward rate constants for the two reactions differ in scale by four orders of magnitude $\left(k_1 / k_2 = 10^{-4}\right)$.

The multiscale nature of the approach is demonstrated by holding $k_1 / k_{-1} = \mathcal{K}_1$ and $k_2 / k_{-2} = \mathcal{K}_2$ constant while varying $k_1 / k_2$ over a large range. Row 3 in Figure 2 shows the steady state concentrations calculated for both coupled reaction theory and stochastic kinetics when $\bar{k}_1/\bar{k}_2$ is varied over eight orders of magnitude ($10^{-4}$ to $10^{4}$) in 100,000 different simulations, each represented as a point in the plot. As the ratio $\bar{k}_1/\bar{k}_2$ increases, the concentration of the reaction intermediates increase as the intermediates are produced faster than they can be taken away until a steady state is reached. Once again, both methods produce exactly the same steady state concentrations.

Steady state concentrations, however, are less sensitive to variations in the rate constants than the rate of material flow through the reactions. The bottom row in Figure 2 shows a similar plot to row 3 except now the rate of material flow through the reaction pathway is plotted as a function of the ratio of the rate constants $\bar{k}_1/\bar{k}_2$. Once again coupled reaction theory and stochastic kinetics produce exactly the same results when the same set of random numbers are used to select which reaction to fire. In fact, the correlation between the trajectories from stochastic kinetics and coupled reaction theory across all values of $\bar{k}_1/\bar{k}_2$ is 1.0 within the numerical precision of the software.

If just one rate constant is available then coupled reaction theory can reproduce the exact time-dependent trajectory as the trajectory governed by the (un-scaled) ordinary differential equation Eqn 19. Measurement of $k_1$ allows one to formulate the time dependence of the reaction intermediates such as $B$ as,

$$\frac{d[B]}{dt} = \tau_1^{-1}\left(1 - \mathcal{K}_{-1}\frac{[B]}{[A]}\right) - c_{21}\left(1 - \mathcal{K}_{-2}\frac{[C]}{[B]}\right). \qquad 24$$

When using Eqn 20, the non-equilibrium transients are accelerated relative to Eqn 19 according to the time scale of the rescaling reaction. (The general form of Eqn. 24 for series of coupled reactions is given in the supplemental data). If the dynamic equations are rescaled by the reaction with the fastest dynamics or highest probability, then the values of the respective time derivatives for the set of differential equations governing the dynamics of the system is reduced accordingly. As a consequence, simulations converge to the steady state in a fewer number of steps. Regardless, the time-to-steady state will be equivalent when time rescaling is used in this manner since reaching the steady state depends on the time scale of the reaction with the lowest probability or slowest dynamics.

To illustrate, consider a pair of reactions similar to those in Scheme A but catalyzed by a pair of enzymes, as shown in Figure 3. Enzyme 1 ($E_1$) binds substrate A and produces product B, which is in turn bound by enzyme $E_2$ to produce product C. Starting from a highly non-equilibrium state consisting of only substrate A and enzymes $E_1$ and $E_2$, initially the fastest reaction will be the binding of A to $E_1$ to form the complex $E_1$A. The slowest reaction will be that producing the final product, C. Using a kinetic simulation based on ordinary differential equations (Eqn 19) and coupled reaction theory based on rescaling (Eqn 20), the dynamics of the un-scaled and time-scaled system over a one second window are shown in Figure 4A for the first enzymatic reaction and Figure 4B for the second enzymatic reaction.

Although both simulations start from the same set of concentrations, the simulation using coupled reaction theory converges to the steady state much more rapidly. The kinetic simulation spends a considerable amount of time simulating the fast dynamics (Figure 4C), while the simulation using coupled reaction theory effectively accelerates the convergence of the fast dynamics and spends relatively little time below the millisecond time scale. Regardless, the steady state is reached in the coupled reaction theory simulation at approximately the same time point as the kinetic simulation because the dynamics of the slowest reactions are unchanged in the vicinity of the steady state.

In addition to being able to produce the correct dynamics without the need for rate constants, coupled reaction theory has an advantage over kinetic formulations of reaction dynamics for multiscale modeling of the steady state: course-graining the dynamics such that the equations are "telescopic" is relatively easy. One can zoom in or out of the details of the reaction system as needed, which can be a considerable advantage for modeling (7). Since coupled reaction theory is based on state functions, it is not necessary to model elementary reactions; summary reactions can be modeled instead and still produce the correct steady state dynamics. The course-grained dynamics will be the composite dynamics of the collapsed system, however. That is, non-steady state transients will not reproduce the detailed dynamics as if all elementary reactions were modeled. Nevertheless, one can always model the detailed dynamics of elementary steps if that is required to address a particular issue.

To demonstrate, one only needs to consider a Kirchoff's loop relationship for cycles (23). Consider the first reaction cycle shown in Figure 3 from reactant A through reactions 2, 3 and 4 to product B and then through reaction -1 back to initial reactant A. The free energy change for traversing around a cycle is zero. As mentioned above, in the deterministic limit (large number of particles) and at steady state, the relation between rate and free energy is $\Delta \mathcal{G} = -RT\log(r_+/r_-)$ where $r_+$ is the forward reaction rate and $r_-$ is the reverse reaction rate (14, 24, 25). For such a cycle at steady state,

$$\frac{r_{+1}}{r_{-1}} = \frac{r_{+2}r_{+3}r_{+4}}{r_{-2}r_{-3}r_{-4}}. \qquad 25$$

For an enzyme catalyzed reaction of a single substrate, $r_{+2}$, $r_{+3}$ and $r_{+4}$ correspond to the rates of binding of the substrate, the catalytic conversion of the substrate to product, and the rate of release of product, respectively. The ratio $r_{+1}/r_{-1}$ then corresponds to the ratio of rates of the combined these steps. This is demonstrated in Figure 4D, which shows the rate ratio of each of the reactions in Figure 3 as the system approaches steady state as well as the product of rate ratios in Eqn 25. Since the course-grained summary reaction 1 does not model the dynamics of enzyme binding, the rate ratio converges to the steady state value rapidly.

That the rate ratios can be predicted using course-grained summary reactions and without the need for rate constants can be a considerable advantage for modeling biological systems. To model the simple two-substrate, two-product enzymatic reaction for the conversion of dihydrofolate to tetrahydrofolate, Fierke et al demonstrated that the kinetic scheme involves 13 reactions (26). In coupled reaction theory, each pair of coupled reactions has the same canonical form shown in Eqn. 20. One can model either the elementary reactions or a summary reaction that describes the overall phenomena as long as the appropriate steady state concentrations are available.

Using coupled reaction theory, measurement of the steady state levels of metabolites and proteins as they exist unbound in the cytosol can be used to derive information on the dynamics of the individual and overall steps of the enzyme-catalyzed reactions, including rate constants (supplementary data). Studies of steady state metabolite levels to-date have focused on the steady state of the cell population under constant growth conditions and not strictly the steady state of the internal metabolites under non-growth conditions. As far as we are aware, comprehensive steady state measurements on metabolites of non-growing cells have not yet been made.

Steady state concentrations alone are not sufficient to derive the dynamics of branched reactions, however. Consider the branched reaction where a product of the first reaction $A \rightleftharpoons B$ is then the reactant for two reactions occurring in parallel,

$$A \underset{k_{-1}}{\overset{k_1}{\rightleftharpoons}} B \begin{array}{l} \underset{k_{-2}}{\overset{k_2}{\rightleftharpoons}} C \\ \underset{k_{-3}}{\overset{k_3}{\rightleftharpoons}} D \end{array}$$

When the concentrations of each species is large enough, the deterministic rate law for species B is,

$$\frac{dB}{dt} = k_1[A] - k_{-1}[B] - k_2[B] + k_{-2}[C] - k_3[B] + k_{-3}[D]$$

Using the approach described above and solving for rate ratios at steady state results in two unknown coupling ratios and one equation, which cannot be solved without additional information. However, if the flux $r_{1,net} = r_1 - r_{-1}$ and $r_{3,net} = r_3 - r_{-3}$ through reactions 1 and 3, respectively, are measured using isotopic labeling assays (metabolic flux analysis (27)), then this additional information allows for determination of the necessary coupling parameters,

$$\frac{r_{3,net}}{r_{1,net}} = \frac{k_3[B] - k_{-3}[D]}{k_1[A] - k_{-1}[B]},$$

or,

$$\frac{k_3[B]}{k_1[A]} = \frac{r_{3,net}}{r_{1,net}} \frac{\left(1 - \mathcal{K}_{-1}\frac{[B]}{[A]}\right)}{\left(1 - \mathcal{K}_{-3}\frac{[D]}{[B]}\right)}$$

In fact, since all reactions in a non-branching pathway all have the same net flux at steady state, *any* steady state flux through a reaction in a non-branching portion of a pathway containing reaction 1 and any steady state flux through a non-branching pathway containing reaction 3 are all that are needed to determine the coupling between reactions 1 and 3.

What can be done if steady state measurements of metabolite levels or reaction fluxes for branched reactions are incomplete or not available? To answer this question, it is necessary to know how the coupling terms (e.g., $c_{21}$ in Eqns 20 and 24) affect the reaction dynamics – in particular, the net rate through a reaction (flux) and the energetic cost to the cell. As can be seen in Figure 2D, changing the ratio of the rate constants $\mathcal{k}_1$ and $\mathcal{k}_2$ can dramatically affect the net rate of the reactions. In particular, a higher ratio $\mathcal{k}_1 / \mathcal{k}_2$ does not lead to a decrease in overall rate; instead the optimum value appears to be around 1. Figure 5A demonstrates the relationship between the coupling term $c_{21}$, the net reaction rate and the total energy of the system. The coupling term is plotted in units of the ratio of the values $\pi(2|J)/\pi(-2|N)/\pi(1|J)/\pi(-1|K)$ where $J$ is the reactant state and $K$ and $N$ are product states of reactions 1 and 2, respectively. There is a many-to-one relationship between $c_{21}$ and the rate because each value of $c_{21}$ is a ratio and can corresponded to multiple combinations of $\bar{k}_1$, $\bar{k}_2$ and the counts/concentrations of each chemical species. Figure 5A is derived from the data in Figure 2C and is the (total) free energy surface as a function of the coupling term $c_{21}$, the ratio $\bar{k}_1 / \bar{k}_2$, and the net reaction rate. The plot is colored by the total free energy of the system at each point.

The average value 1.0 for the coupling term $c_{21}$ has particular significance – it is the maximum entropy steady state out of all the possible steady states. The maximum entropy steady state is the thermodynamically most efficient steady state (least heat dissipation) and is characterized by being the steady state of lowest total free energy and highest likelihood. As shown in Figure 5A for a non-equilibrium driving force of 2 kcal/mol, it is possible to maximize the net rate but at the cost of thermodynamic efficiency. That is, a somewhat higher net rate can be achieved, but the system must be maintained at a higher energy. However, at even moderate thermodynamic driving forces of 5 kcal/mol for the combined reactions, there is no speed advantage – the net rate is already maximized at the thermodynamic optimum. As long as both reactions individually have moderate non-equilibrium driving forces, there is no speed advantage to tweaking rate constants. Rate and thermodynamic efficiency are meaningfully combined in the concept of entropy production rate, hinted at by Lotka (28, 29) and proposed by Schrödinger and many others since (30). The definition of entropy production rate used here is the product of the net rate and the entropy change in going from initial products to final reactants (19),

$$r_{net} \cdot S = -r_{net} \cdot \sum_{i}^{\text{reactions}} \sum_{J}^{\text{states}} \pi_i(J) \log \pi_i(J)$$

Figure 5B shows that there is a considerable decrease in steady state entropy production rates as one moves away from the thermodynamically optimal state. If one assumes that adaptation favors organisms with greater entropy production, then one can create reasonable models without even the need for steady state values of reaction intermediates by setting the coupling term to the thermodynamic odds ratio of the coupled reactions. Already, a correlation between the standard entropy changes of overall reaction pathways and their activities has been demonstrated (31). If metabolite measurements are available, the hypothesis of maximum entropy for steady states of evolutionary optimized systems is testable (32, 33). It is important to note however, that while entropy production may be maximized at steady state, steady state entropy production rates are constrained by a fluctuation theorem that is a function of the rates achievable by respective enzymes (34).

From an evolutionary perspective, conservation of a specific rate constant would be very hard and would require strict error tolerances in replication, which would likely lead to decreased ability to adapt. More likely, fitness fluctuates around the state of maximum entropy production. However, it must be kept in mind that the coupled reactions of metabolism do not represent the total thermodynamics of the cell and it is even possible for individual reaction entropy to decrease rather than increase. Feedback regulation of enzyme activities may occur for related reasons, such as regulating the production of metabolites to synchronize the cell's network of coupled reactions with those of the environment (35), or entirely different reasons. Regardless, predictive models of complex adaptive systems such as those found in biology and elsewhere that do not depend on hard-to-measure parameters are urgently needed to accelerate research in systems relevant to medicine, climate and energy challenges. We have outlined a statistical thermodynamic framework that would enable such large-scale simulations. The approach does not rely on kinetic parameters, but rather on standard free energy values and metabolite concentrations for which robust measurement methods are being developed (36).

**Acknowledgements:** This work was supported by the U.S. Department of Energy's Office of Biological and Environmental Research and the Environmental Molecular Sciences Laboratory (EMSL) and. EMSL is a national scientific user facility operated by PNNL for the Office of Biological and Environmental Research at the U. S. Department of Energy. PNNL is operated by Battelle for the U.S. Department of Energy under Contract DE-AC06-76RLO.

**Figure Legends**

Figure 1. Flux through a state *J*. Each state *J, K, L, M,* and *N* is defined by the sets of counts or concentrations of the chemical species of the system. Due to the four reactions 1, -1, 2 and -2, only four other states are adjacent to state *J*: *K, L, M,* and *N*. In the non-equilibrium process the forward reactions 1 and 2 are much more likely than the reverse reactions -1 and -2.

Figure 2. (A - D) Comparisons between stochastic kinetic simulations and simulations using coupled reaction theory in which the same set of random numbers were used in both sets of simulations. Simulations of the coupled reactions of Scheme A with $\gamma_A = \gamma_B = \gamma_C = 1$ were carried out at different driving forces by fixing C and setting the boundary concentrations of A to an appropriate level. In each case, the equilibrium constant for reaction 1 ($K_1$) is 25-fold greater than that for reaction 2 ($K_2$). In both (A) and (B) the coupled reaction theory trajectory is offset by +5 counts so that it can be distinguished. (A) Steady state trajectory from a typical simulation. (B) Non-equilibrium transient trajectories from a typical simulation. (C) Steady state counts of the intermediate *B* from 100,000 simulations in which the ratio of the rate constants $k_1 / k_2$ is varied from $10^{-4}$ to $10^4$ while keeping $k_1 / k_{-1} = K_1$ and $k_2 / k_{-2} = K_2$. (D) Steady state net reaction rate (flux) values over the same set of simulations as in (C).

Figure 3. Coupled enzyme catalyzed reactions corresponding to the uncatalyzed reactions of Scheme A. The uncatalyzed reactions are 1 and 5. In the catalyzed reactions enzyme 1 binds substrate A in reaction 2, converting the substrate to intermediate B in reaction 3, and releasing B in reaction 4, while enzyme 2 binds the intermediate B in reaction 6, converts the intermediate to product C in reaction 7, and releases the product in reaction 8.

Figure 4. Time dependence of enzymatic reaction 1 (A) and reaction 2 (B) using rate parameters (labeled 'kinetic'), coupled reaction theory with time-rescaling (labeled CRT) and coupled reaction theory of summary reactions that do not include catalyst dynamics (labeled CRT). The latter can be thought of as the statistical course graining of the reaction dynamics of the detailed system in which the summary reaction represents the course grained system. In both coupled reaction theory simulations the processes occurring on small time scales (fast dynamics) converge much faster than in the simulation based on rate parameters. (C) Plot of the time scale of the dynamics as a function of the simulation step. The time rescaling results in larger time steps taken per simulation step for both CRT simulations. (D) The ratio of the rates for each of the reactions in Figure 3. Reactions 2-4 and 6-8 involve enzyme dynamics while reactions 1 and 4 are summary reactions representing the overall process. At steady state, a Kirchhoff loop law is obeyed such that Eqn 25 holds. All subscripts refer to the reaction scheme shown in Figure 3. Reaction parameters for Figure 4 are as follows. Equilibrium constants: $K_1 K_5 = 100$, $K_1 = 0.2$, $K_5 = 500$, $K_2 = 10^6$, $K_3 = 0.2$, $K_4 = 10^{-6}$, $K_6 = 10^6$, $K_7 = 500$, and $K_8 = 10^{-6}$. Rate parameters: $k_2 = 10^9$, $k_{-2} = 10^3$, $k_3 = 10^3$, $k_{-3} = 5 \cdot 10^3$, $k_4 = 10^3$, and $k_{-4} = 10^9$, $k_6 = 10^9$, $k_{-6} = 10^3$, $k_7 = 10^3$, $k_{-7} = 2$, $k_8 = 10^3$, $k_{-8} = 10^9$.Concentrations: Figure 4A-4C, $[A] = 10^{-3}$ and $[C] = 1.4373 \cdot 10^{-25}$ (both fixed); For Figure 4D, the initial condition is $[C] = 1.4373 \cdot 10^{-9}$ and the initial concentrations for enzymes is $[E_1] = 0.5*10^{-3}$ and $[E_2] = 0.5*10^{-3}$.

Figure 5. Each colored region represents different combinations of rate constants, varying from $k_1 / k_2$ = $10^{-4}$ to $10^4$, as in Figure 2C-D. (A) Comparison of net rate (flux), $r_{net}$ and total free energy (color) as a function of the coupling constant $c_{21} = \pi_2 / \pi_1$ for the reaction system described in Scheme A and caption 1. The coupling constant is given in units of the thermodynamic likelihood $\pi_2 / \pi_{-2} / \pi_1 / \pi_{-1}$ . If the two reactions are equally likely from a thermodynamic perspective, then their likelihood ratio is 1.0. (B) Entropy production rate as a function of the coupling constant, $c_{21}$ . The entropy production rate is near maximum (driving forces close to equilibrium) or maximized (far from equilibrium) when $\pi_2 / \pi_{-2} / \pi_1 / \pi_{-1} = 1$.

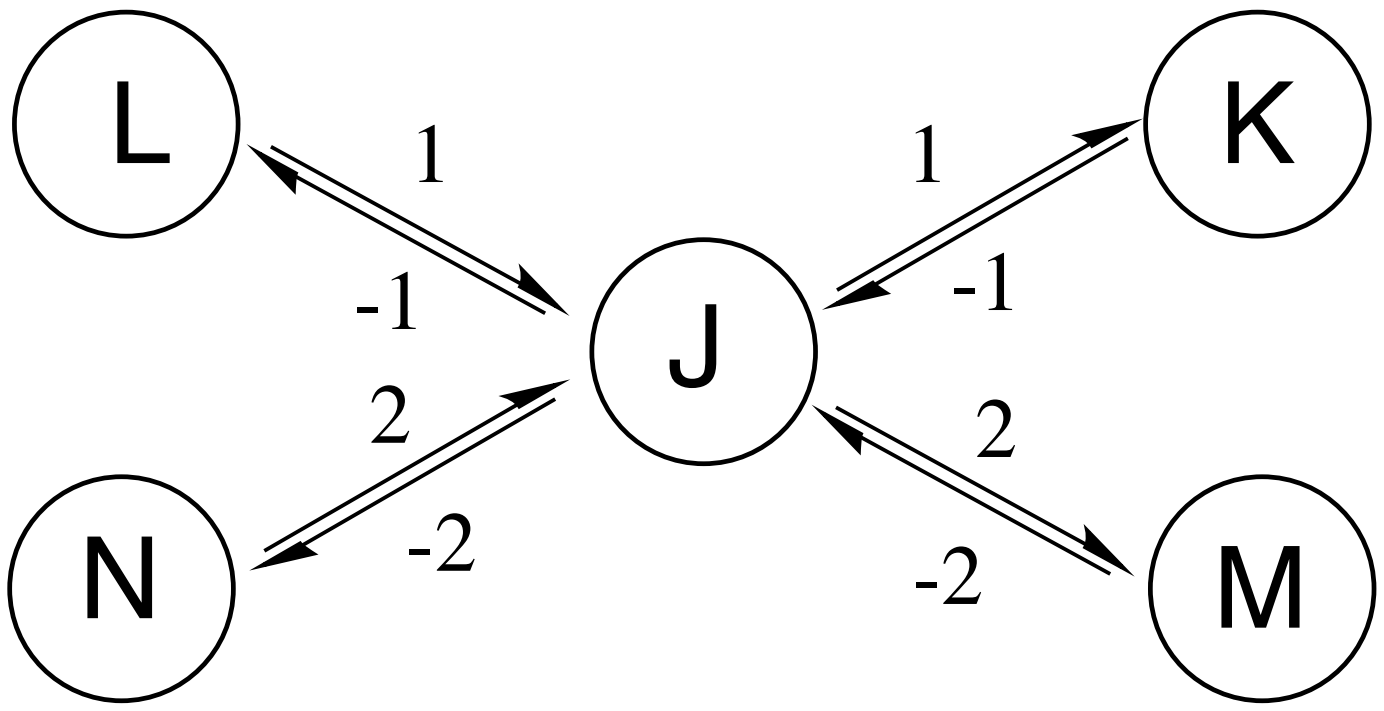

5

Figure 1.

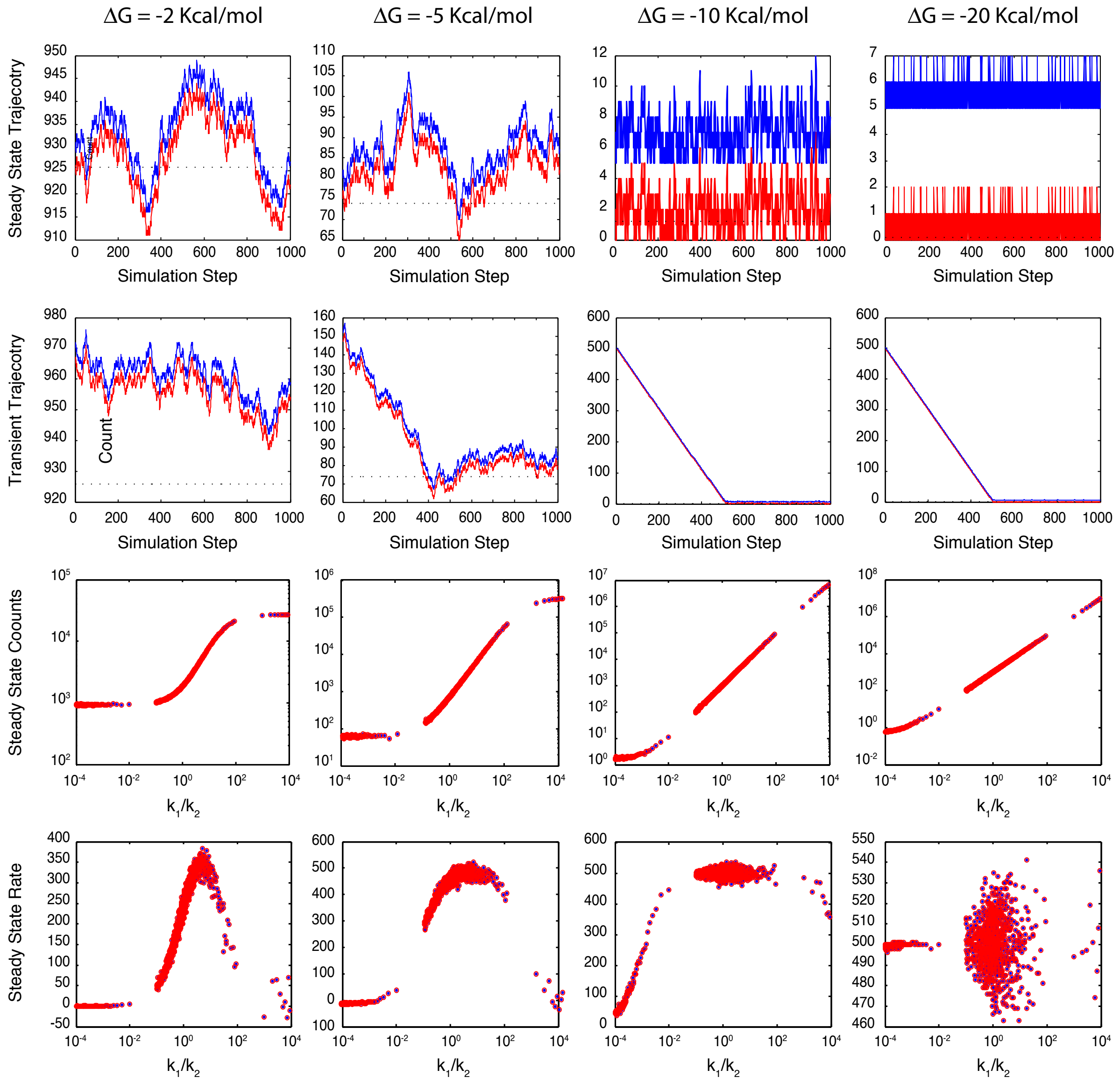

ΔG = -2 Kcal/mol
ΔG = -5 Kcal/mol
ΔG = -10 Kcal/mol
ΔG = -20 Kcal/mol
Steady State Trajecotry
Transient Trajecotry
Steady State Coounts
Steady State Rate
Simulation Step
Count
k1/k2

Figure 2.

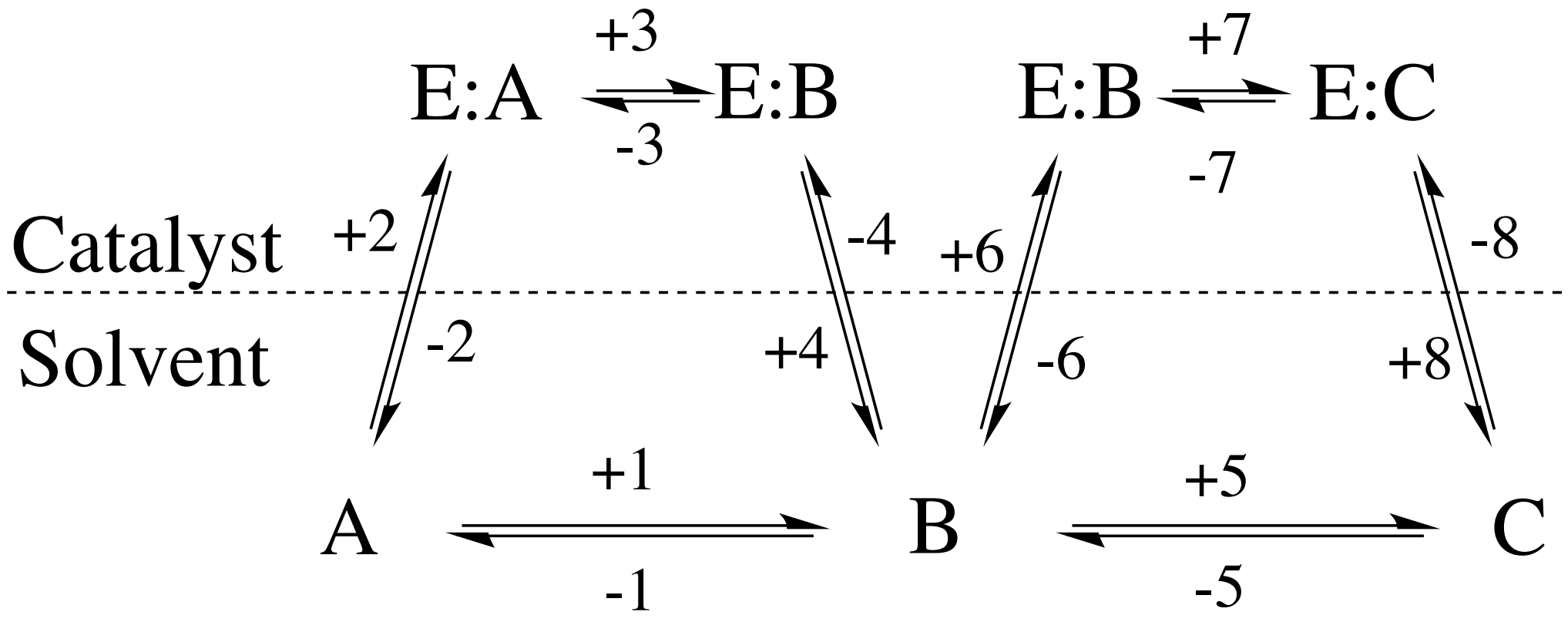

Figure 3.

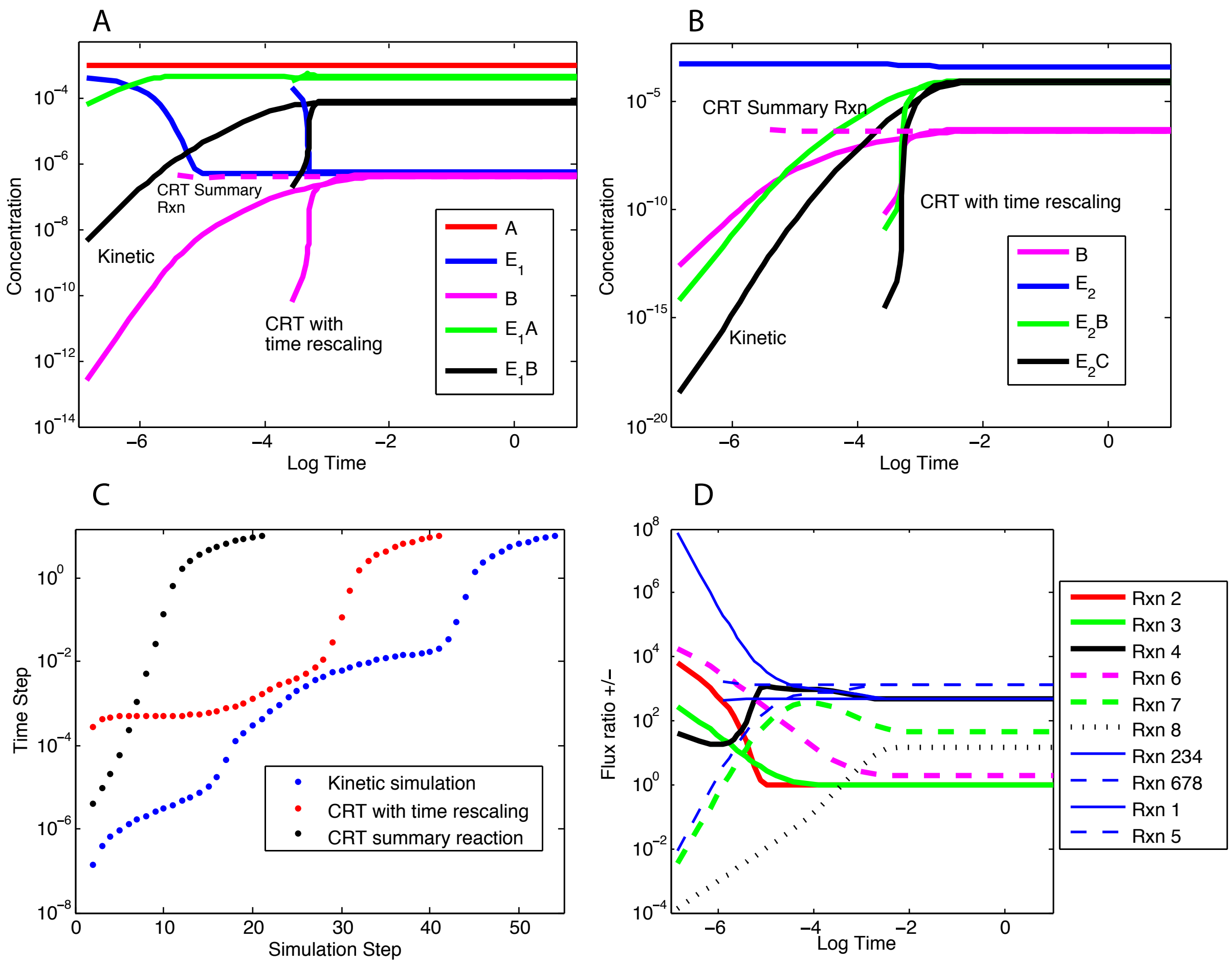

Figure 4.

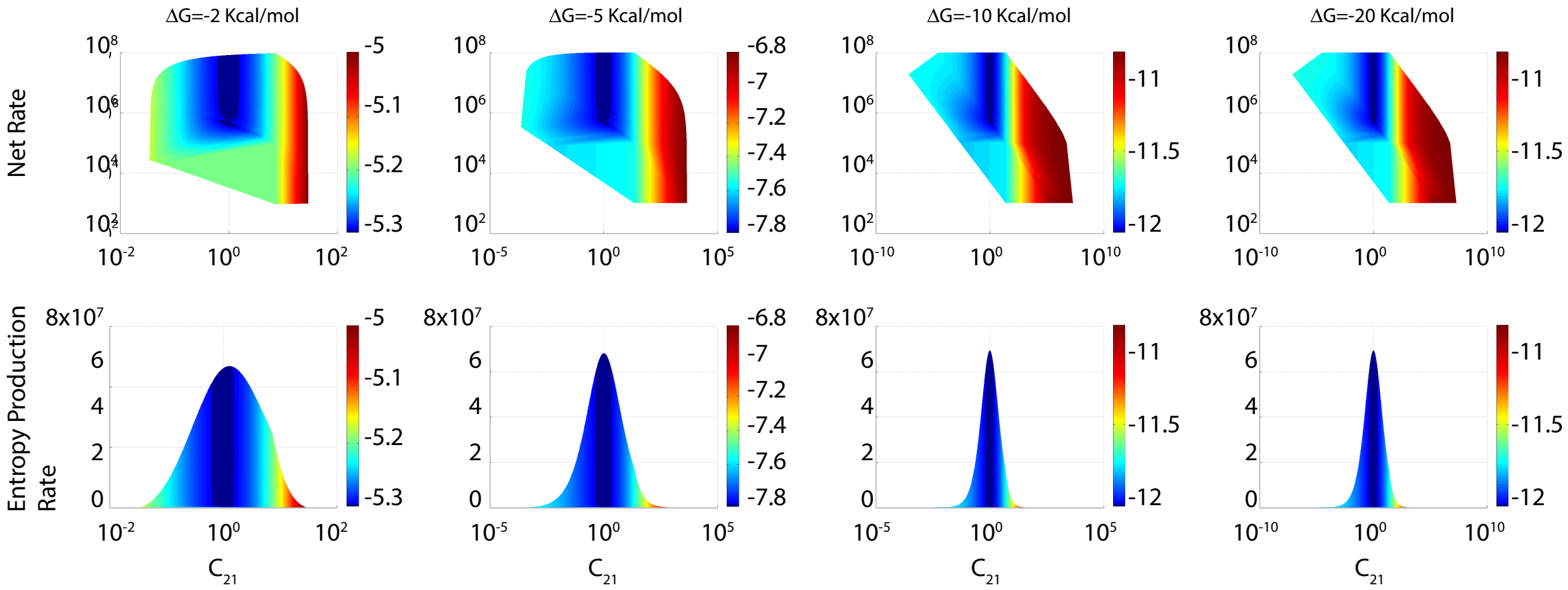

Figure 5.